\documentclass{PoS}

\usepackage{amsmath}
\usepackage{amssymb}

\newcommand{\be}{\begin{equation}}
\newcommand{\ee}{\end{equation}}
\newcommand{\beq}{\begin{eqnarray}}
\newcommand{\eeq}{\end{eqnarray}}

\title{   \begin{picture}(0,0)(0,0)%
   \put(355,75){\makebox(0,0)[l]{\textnormal{\normalsize UTHEP-592}}}%
   \end{picture}%
   Calculation of nucleon strange quark content  with dynamical overlap quarks}

\ShortTitle{nucleon strange quark content with dynamical overlap quarks}

\author{
JLQCD collaboration:\,\, 
\speaker{K. Takeda}$^a$, 
S.~Aoki$^{a,b}$, 
S.~Hashimoto$^{c,d}$, 
T.~Kaneko$^{c,d}$,
T.~Onogi$^e$, 
N.~Yamada$^{c,d}$ 
\\
\llap{$^a$}Graduate School of Pure and Applied Sciences,
           University of Tsukuba, Tsukuba, Ibaraki, 305-8571, Japan\\
\llap{$^b$}Riken BNL Research Center, Brookhaven National Laboratory,
           Upton, New York 11973, USA\\
\llap{$^c$}High Energy Accelerator Research Organization (KEK), 
           Ibaraki 305-0801, Japan\\
\llap{$^d$}School of High Energy Accelerator Science, 
           The Graduate University for Advanced Studies\\
           (Sokendai), Ibaraki 305-0801, Japan\\
\llap{$^e$}Department of Physics, Osaka University Toyonaka, 
           Osaka 560-0043, Japan\\
E-mail: \email{ktakeda@het.ph.tsukuba.ac.jp}
}

\abstract{
We calculate the nucleon strange quark content  
directly from disconnected three-point functions. 
Numerical simulations are carried out in two-flavor QCD 
using the overlap quark action with up and down quark masses 
down to a fifth of the physical strange quark mass. 
To improve the statistical accuracy,
we calculate the nucleon two-point functions 
with the low-mode averaging technique, 
whereas 
the all-to-all quark propagator is used for the disconnected quark loop.
We obtain the $y$ parameter,
which is the ratio of the strange and light quark contents, 
$y = 0.024(45)$ at the physical point. 
This is in a good agreement with our earlier calculation 
from the nucleon spectrum through the Feynman-Hellmann theorem.
}

\FullConference{The XXVII International Symposium on Lattice Field Theory - LAT2009\\
		 July 26-31 2009\\
		 Peking University, Beijing, China}

\begin{document}


\section{Introduction}

The nucleon strange quark content $\langle N|\bar{s}s|N\rangle$ 
is an important parameter to determine the cross section of the scattering 
of dark matter candidates from the nucleon \cite{DM1,DM2}.
It can not be measured directly by experiments, 
and only lattice QCD can provide
a model-independent and nonperturbative determination.
A precise lattice calculation is, however, very challenging, 
because 
only disconnected diagrams contribute to the strange quark content
and they are computationally very expensive to calculate
with the conventional method.
In addition,
the scalar operator $\bar{s}s$ has a vacuum expectation value (VEV),
which diverges towards the continuum limit.
We need to subtract the VEV contribution 
and this induces a substantial uncertainty in the strange quark content.

In our previous study \cite{ohki},
we avoid the above mentioned difficulties
by calculating the strange quark content from the quark mass dependence of the nucleon mass 
$m_N$ through the Feynman-Hellmann theorem
\be
   \langle N | \bar{s}s | N \rangle
   = 
   \frac{\partial m_N}{\partial m_s}.
   \label{eqn:FH}
\ee
We refer to this method as the spectrum method in the following.
This method is, however, not applicable to other interesting matrix elements,
such as the strange quark spin fraction of the nucleon.
In this article, therefore, 
we attempt a direct determination of strange quark content 
from nucleon matrix element including a disconnected diagram.
To this end, 
we employ the overlap quark action,
which has exact chiral symmetry,
and improved measurement methods, 
such as the low-mode averaging (LMA) technique \cite{LMA1,LMA2}
and the use of the all-to-all quark propagator \cite{ata}.

\section{Simulation details}

Gauge ensembles of two-flavor QCD are generated 
on a $L^3 \times T= 16^3 \times 32$ lattice
using the Iwasaki gauge action and the overlap quark action.
We set the gauge coupling $\beta=2.30$ at which 
the lattice spacing determined from the Sommer scale $r_0=0.49\,$fm
is $a=0.118(2)\,$fm. 
Our simulation is accelerated 
by introducing a topology fixing term into our lattice action \cite{fixedQ},
and we simulate only the trivial topological sector $Q=0$ 
in this study.
We take four values of bare up and down quark masses
$m_{ud}=0.015,0.025,0.035$ and 0.050,
which cover a range of the pion mass $m_{\pi}=290-520$~MeV. 
Statistics are 100 independent configurations at each quark mass. 
We refer readers to ~\cite{Nf2:JLQCD} 
for further details on our configuration generation.
In our measurement,
we take two values of the valence strange quark mass 
$m_{s,val}=0.070$ and 0.100,
which are close to the physical mass $m_{s,phys}=0.077$ 
determined from our analysis of the meson spectrum ~\cite{m_s}.


The strange quark content can be extracted 
from nucleon two- and three-point functions
\begin{eqnarray}
C_{2pt}^\Gamma(t,\Delta t) 
& = & 
\mathrm {Tr}[ \Gamma \langle N(t+\Delta t)\bar{N}(t)\rangle], 
\label{2pt}
\\
C_{3pt}^\Gamma(t,\Delta t,\Delta t_s)
& = &
\mathrm{Tr} [\Gamma \langle N(t+\Delta t)S(t+\Delta t_s)\bar{N}(t)\rangle]
-\langle S(t+\Delta t_s)\rangle\mathrm 
 {Tr}[ \Gamma \langle N(t+\Delta t)\bar{N}(t)\rangle], 
\label{3pt}
\end{eqnarray}
where $S=\bar{s}s$ is the strange scalar operator,
$t$ represents the temporal coordinate of the nucleon source operator,
and $\Delta t$ ($\Delta t_s$) is 
the temporal separation between the nucleon source and sink (quark loop).
We calculate $C_{2pt}^\Gamma$ and $C_{3pt}^\Gamma$ 
with two choices of the projector $\Gamma=\Gamma_\pm = (1\pm\gamma_4)/2$
corresponding to the forward and backward propagation of the nucleon.
We then take the average over the two choices of $\Gamma$ with 
appropriately chosen temporal separations $\Delta t$ and $\Delta t_s$.
The averaged correlators, which are denoted by $C_{2pt}$ and $C_{3pt}$ 
in the following, show reduced statistical fluctuation.


For further improvement of the statistical accuracy,
we employ the low-mode averaging(LMA) technique \cite{LMA1,LMA2} to calculate $C_{2pt}$ and $C_{3pt}$.
In this method, 
the quark propagator is expanded 
in terms of the eigenmodes of the Dirac operator $D$.
We calculate the contribution of 100 low-modes exactly as 
\be 
 (D)_{low}^{-1}=\sum_{i=1}^{100}\frac{1}{\lambda^{(i)}}v^{(i)}v^{(i) \dagger},
 \hspace{3mm}
 \qquad D\,v^{(i)}=\lambda^{(i)} v^{(i)}.
 \label{DovL}
\ee
The remaining contribution from the higher modes is taken from
by that of the conventional point-to-all propagator.
With this decomposition of the quark propagator, 
$C_{2pt}$ is divided into eight contributions 
\beq
   C_{2pt}
   = 
   C_{2pt}^{LLL}
  +C_{2pt}^{LLH}+C_{2pt}^{LHL}+C_{2pt}^{HLL}
  +C_{2pt}^{LHH}+C_{2pt}^{HLH}+C_{2pt}^{HHL}
  +C_{2pt}^{HHH}.
  \label{C2pt:cntrb}
\eeq
It is expected that $C_{2pt}^{LLL}$ dominates 
$C_{2pt}$ at large temporal separation $\Delta t$.
The statistical accuracy of $C_{2pt}$ can be remarkably improved
by averaging $C_{2pt}^{LLL}$ over the location of the nucleon source operator.
We also improve the statistical accuracy of other contributions 
($C_{2pt}^{LLH}$, ..., $C_{2pt}^{HHH}$)
by using point-to-all propagators
averaged over 4 or 8 different source locations.
The nucleon piece of the disconnected correlator $C_{3pt}$ 
is calculated in the same way.


Since the nucleon correlators $C_{2pt}$ and $C_{3pt}$ 
damp rapidly as $\Delta t$ increases, 
it is essential to reduce the contamination from excited states at small 
$\Delta t$. 
In this study, we employ the Gaussian smearing 
\be
q_{smr}({\bf x},t) 
= \sum_{\bf y} \left\{ \left( 1+\frac{\omega}{4N} \,\,  H \right)^N \right\}_{{\bf x,y}} 
             q({\bf y},t), \label{Gss_fn} \qquad
H_{{\bf x,y}} 
= \sum_{i=1}^3 (\delta_{{\bf x,y}-i}+\delta_{{\bf x,y}+i})
\ee
for both of the source and sink operators. 
The parameters $\omega=20$ and $N=400$ are chosen 
so that the effective mass of $C_{2pt}$ shows a good plateau.
For comparison, we repeat our measurement with the local sink operator.
In this additional measurement, 
we test the local and an exponential source operator
$q_{smr}({\bf x},t) = \sum_{\bf r}\exp(-B|{\bf r} |) q({\bf {x+r}},t)$.
The parameter $B$ is chosen so that 
the distribution of the smeared quark is close to that of the 
Gaussian smearing (\ref{Gss_fn}).


To calculate the disconnected quark loop in $C_{3pt}$,
we construct the all-to-all quark propagator as proposed in ~\cite{ata}.
The low-mode contribution is the same as in (\ref{DovL})
and the contribution from the high-modes is estimated 
by employing the noise method with the dilution technique \cite{ata}. 
We prepare a single $Z_2$ noise vector $\eta$ for each configuration 
and it is split into $N_{d}=3\times 4 \times T/2$ vectors 
$\eta^{(d)}$ ($d=1,...,N_d$)
which have non-zero elements only for single color and spinor indices and 
two consecutive time-slices. 
The high-mode contribution is then given by
\be 
(D)_{high}^{-1}
=
\sum_{d=1}^{N_{d}}\psi^{(d)}\eta^{(d) \dagger}, 
\label{DovH}
\ee
where $\psi^{(d)}$ is the solution of the linear equation
\be 
D\,\psi^{(d)}=(1-\mathcal P_{low})\eta^{(d)}
\ee
and $\mathcal P_{low}$ is the projection operator 
to the subspace spanned by the low-modes. 

We also tested the all-to-all quark propagator
to calculate the high-mode contributions $C_{2pt}^{LLH}, ..., C_{2pt}^{HHH}$
in ~(\ref{C2pt:cntrb}).
It turned out, however, that 
these contributions have large statistical error due to 
the insufficient number of the noise samples.
We therefore use $C_{2pt}$ and $C_{3pt}$ calculated with the LMA
in the following analysis.

\section{Matrix element at simulated quark masses} 
\label{Sec.3}

\begin{figure}[t]
  \centering
  \includegraphics[width=0.9\textwidth,clip]{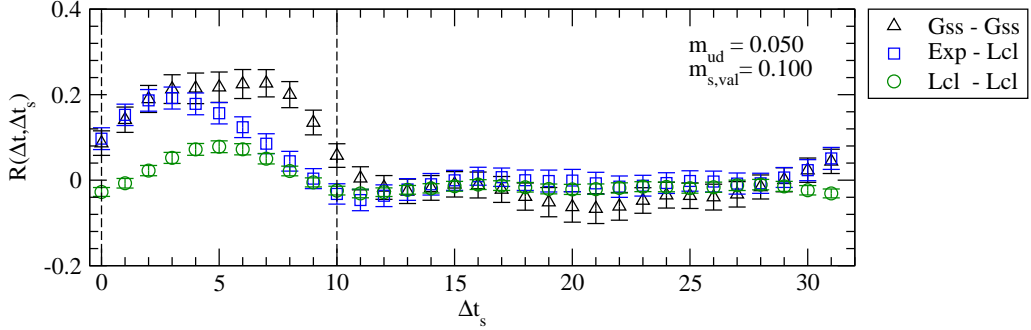} 
  \vspace{-3mm}
  \caption{
    Ratio $R(\Delta t,\Delta t_s)$ at $m_{ud}=0.050$ and $m_{s,val}=0.100$
    with $\Delta t$ fixed to 10.
    Triangles show data with Gaussian smeared source and sink,
    whereas circles (squares) are with the local sink and 
    local (exponentially smeared) source.
    We omit the noisy high-mode contribution to the quark loop 
    in this plot.
    Vertical lines show the location of the source and sink operators.
  }
  \label{fig:comp_smr}
\end{figure}

\begin{figure}[b]
  \centering
    \includegraphics[width=0.9\textwidth,clip]{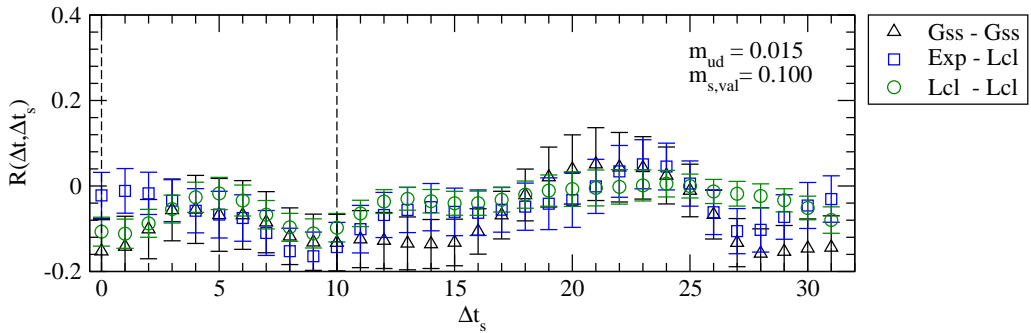}
    \vspace{-3mm}
    \caption{
      Ratio $R(\Delta t,\Delta t_s)$ at $m_{ud}=0.015$ and $m_{s,val}=0.100$.
    }
    \label{fig:light_mass}
\end{figure}

We extract the unrenormalized matrix element 
$\langle N|\bar ss|N\rangle_{lat}$ from the ratio
\beq
   R(\Delta t,\Delta t_s) 
   = \frac{C_{3pt}(t,\Delta t,\Delta t_s)}{C_{2pt}(t,\Delta t)}
   \xrightarrow[\Delta t, \Delta t_s \to \infty]{} 
   \langle N|\bar ss|N\rangle_{lat}.
\eeq
Figure~\ref{fig:comp_smr} shows $\Delta t_s$ dependence of 
$R(\Delta t, \Delta t_s)$ at our heaviest $ud$ quark mass $m_{ud}\!=\!0.050$ 
with a fixed $\Delta t$.
We observe a clear plateau between the nucleon source and sink
with the Gaussian smeared operator.
On the other hand, the plateau is unclear 
if the local operator is used for the source and/or sink.
It is therefore crucial 
for a reliable determination of 
$\langle N|\bar ss|N\rangle_{lat}$ from $R(\Delta t,\Delta t_s)$
to reduce contamination from the excited states 
by appropriately smearing the nucleon operators.

The situation is similar 
at two smaller $ud$ quark masses $m_{ud}\!=\!0.035$ and 0.025.
As shown in Fig.~\ref{fig:light_mass},
however,
we do not observe 
a clear signal even with the smeared source and sink
at our smallest quark mass $m_{ud}\!=\!0.015$.
To observe a clear plateau of $R(\Delta t,\Delta t_s)$
at such small $m_{ud}$, 
we may need more statistics as well as a larger lattice
to suppress finite volume corrections, which are possibly sizable
at $M_\pi L \sim 2.8$ at $m_{ud}=0.015$.
We leave such a calculation for future studies,
and omit data at this $m_{ud}$ in the following analysis.
  
\begin{figure}[t]
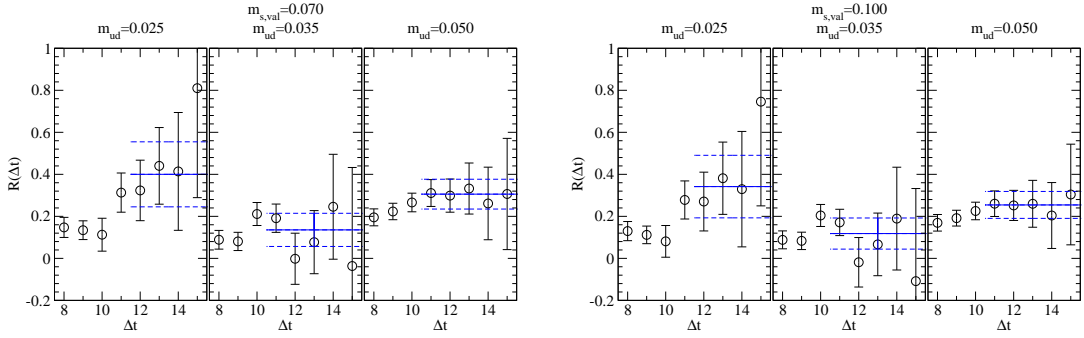

  \begin{tabular}{cc}
    \centering
    \begin{minipage}{6.8cm}
      \includegraphics[width=6.8cm,clip]{fit_m070.eps}
    \end{minipage}
    \hspace{5mm}
    \begin{minipage}{6.8cm}
      \includegraphics[width=6.8cm,clip]{fit_m100.eps}
  \end{minipage}
  \end{tabular}
  \caption{
    Ratio $R(\Delta t)$ as a function of $\Delta t$.
    Three left (right) panels show data at $m_{s,val}=0.070$ (0.100).
    The horizontal lines show 
    the constant fit to $R(\Delta t)$ in terms of $\Delta t$
    to determine $\langle N|\bar ss|N\rangle_{lat}$.
  }
  \label{fig:fit_dt}
\end{figure}

In this report,
$\langle N|\bar ss|N\rangle_{lat}$
is determined by the following simple two-step fits.
First, we carry out a constant fit to $R(\Delta t,\Delta t_s)$ 
in terms of $\Delta t_s$.
The fit result, which we denote by $R(\Delta t)$, is plotted 
as a function of $\Delta t$ in Fig.~\ref{fig:fit_dt}.
We then extract $\langle N|\bar ss|N\rangle_{lat}$
by a constant fit to $R(\Delta t)$ at $12 \leq \Delta t \leq 15$.
As seen in Fig.~\ref{fig:fit_dt},
$R(\Delta t)$ do not show significant $\Delta t$ dependence
with this range of $\Delta t$.
We therefore expect that 
extraction of ground state signal is well under control.

\section{Strange quark content at physical point}
\label{Sec.4}

\begin{figure}[t]
  \centering
    \includegraphics[width=.8\textwidth,clip]{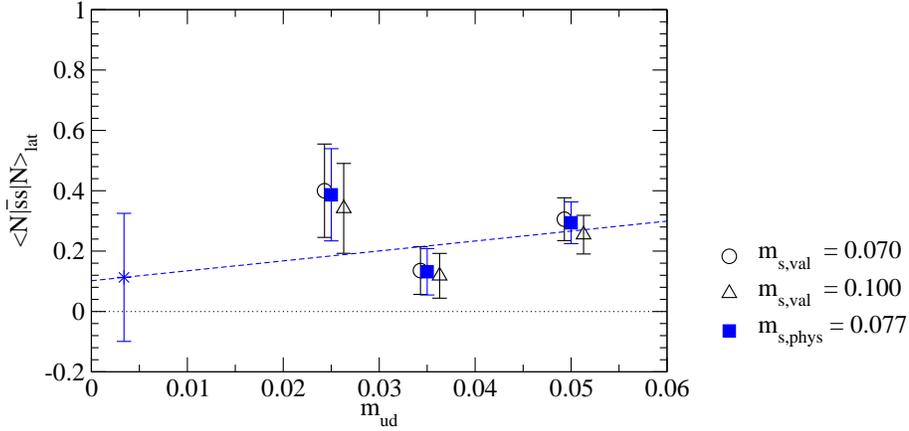}
    \vspace{-2mm}
    \caption{
      Chiral extrapolation of $\langle N | \bar{s}s | N \rangle_{lat}$ 
      at $m_{s,phys}$ as a function of $m_{ud}$.
      For a comparison,
      we also plot $\langle N | \bar{s}s | N \rangle_{lat}$ 
      at simulated strange quark masses $m_{s,val}\!=\!0.070$ and 0.100.
    }
    \label{fig:chiral fits}
\end{figure}

As seen in Fig.~\ref{fig:fit_dt}, 
the fit result for $\langle N|\bar ss|N\rangle_{lat}$
does not have significant $m_{s,val}$ dependence at each $m_{ud}$.
This leads us to interpolate $\langle N|\bar ss|N\rangle_{lat}$
to the physical strange quark mass $m_{s,phys}=0.077$
using a linear form in terms of $m_{s,val}$.
Fit results as well as fitted data at $m_{s,val}$ are plotted 
as a function of $m_{ud}$ in Fig.~\ref{fig:chiral fits}.

At next-to-leading order of heavy baryon chiral perturbation theory (HBChPT) ~\cite{HBChPT} 
the nucleon mass $m_N$ can be written as 
$m_N=m_0 + C_1 m_{ud} + C_2 m_s + C_3 m_\pi^3 + C_4 m_K^3 + C_5 m_\eta^3$,
where $C_i$ $(i=1,...,5)$ are functions of the low-energy constants (LECs) 
in HBChPT.
We note that the contributions from decuplet baryons are neglected.
The Feynman-Hellmann theorem (\ref{eqn:FH}) then implies 
that
the $m_{s}$ dependence of $\langle N | \bar{s}s | N \rangle$ 
comes from the $O(m_{K,\eta}^3)$ terms from $K$ and $\eta$ loops.
By using the leading order relation 
$m_K^2 \propto m_{ud} + m_s$ and $m_\eta^2 \propto m_{ud} + 2 m_s$,
we obtain
\beq
   \langle N | \bar{s}s | N \rangle
   & = &
   \frac{\partial m_N}{\partial m_s}
   = 
   D_0  + D_1 m_{ud} + O(m_{ud}^2),
   \label{eqn:chiral_fit}
\eeq
where $D_0$ and $D_1$ depend on the LECs and $m_s$.

We extrapolate $\langle N | \bar{s}s | N \rangle_{lat}$ at $m_{s,phys}$ 
by the linear form (\ref{eqn:chiral_fit}) 
with $D_{0,1}$ treated as fitting parameters.
This chiral extrapolation is plotted in Fig.~\ref{fig:chiral fits}.
We obtain $\langle N|\bar ss|N\rangle=0.11(21)$ at the physical point,
where the error is statistical only. 
This is converted to the phenomenologically relevant parameters 
\beq
f_{T_s} \equiv \frac{m_{s,phys} \langle N|\bar ss |N \rangle}{m_N}
        = 0.015(28),
\eeq
\vspace{-2mm}
and 
\vspace{-2mm}
\beq
y \equiv \frac{2\langle N|\bar ss|N\rangle}{\langle N|\bar uu+\bar dd|N\rangle}
  = 0.024(45),
\label{eqn:y}  
\eeq
where we use 
the nucleon mass $m_N$ \cite{m_N}
and the $ud$ quark content $\langle N|\bar uu+\bar dd|N\rangle$
obtained in our previous study~\cite{ohki}.

\begin{figure}[h]
  \centering
  \includegraphics[width=.6\textwidth,clip]{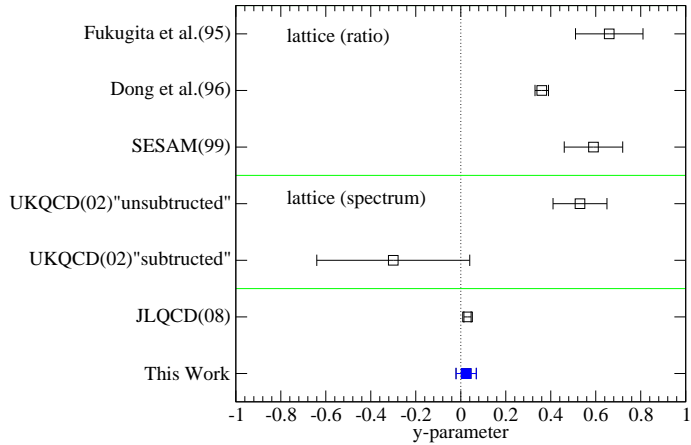}
  \vspace{-2mm}
  \caption{
    Comparison of $y$ parameter with previous estimates
    \cite{ohki,Fukugita,Dong,SESAM,UKQCD}.    
    Our previous study \cite{ohki} employs the overlap action,
    whereas other studies use the Wilson-type quark actions.
    Two results from the UKQCD study \cite{UKQCD} are obtained 
    with and without subtracting the unphysical effect 
    due to the operator mixing.
  }
  \label{fig:comp}
\end{figure}

As shown in Fig.\,\ref{fig:comp}, 
we observe a good agreement 
with our estimate from the spectrum method \cite{ohki}. 
The same figure also shows that previous studies with 
the Wilson-type fermions \cite{Fukugita,Dong,SESAM}
led to rather large values for the strange quark content $y$.
It is argued in ~\cite{ohki,UKQCD}
that the explicit chiral symmetry breaking 
induces a mixing between the scalar operators of sea and valence quarks 
and leads to a substantial uncertainty in the strange quark content.

\section{Conclusion}
\label{Sec.5}

In this article,
we report on our calculation of the strange quark content
directly from the nucleon matrix element.
We determine $f_{T_s}$ and $y$ with an accuracy of $O(10^{-2})$.
The key points leading to this accuracy 
are the use of the improved measurement techniques,
namely the LMA and the all-to-all quark propagator,
as well as the appropriately smeared operator 
both for nucleon source and sink.
It is an interesting subject in the future 
to extend this study to other matrix elements containing disconnected diagram
such as the quark spin fraction of the nucleon. 
 
We observe a good agreement 
with our previous estimate from the spectrum method.
Chiral symmetry preserved by the overlap action 
plays a crucial role in avoiding 
the unwanted operator mixing for the Wilson-type actions.
For more precise determination,
we need to extend our calculation to $N_f=2+1$ QCD and larger volumes.
Our preliminary estimate with the spectrum method is 
reported at this conference \cite{ohki:lat09}.
A direct determination from nucleon disconnected functions 
in $N_f=2+1$ QCD is also in progress.

\vspace{5mm}

Numerical simulations are performed 
on Hitachi SR11000 and IBM System Blue Gene Solution 
at High Energy Accelerator Research Organization (KEK)
under a support of its Large Scale Simulation Program (No.~09-05).
This work is supported in part by the Grant-in-Aid of the
Ministry of Education 
(No.~19540286, 20105001, 20105002, 20105003, 20340047, 21674002 and 21684013).



\begin{thebibliography}{99}

\bibitem{DM1} 
E. A.~Baltz, M.~Battaglia, M.E.~Peskin and T.~Wizansky, 
\emph{Phys. Rev. D} {\bf 74}, 103521 (2006)

\bibitem{DM2} 
J.~Ellis, K.A.~Olive, C.~Savage, 
\emph{Phys. Rev. D} {\bf 77}, 065026 (2008)
[{\tt 0801.3656[hep-ph]}]. 

\bibitem{ohki} 
H.~Ohki {\it et. al}. (JLQCD collaboration), 
\emph{Phys. Rev. D} {\bf 78}, 054502 (2008)
[{\tt 0806.4744[hep-lat]}].

\bibitem{LMA1}
L.~Giusti, P.~Hernandez, M.~Laine, P.~Weisz and H.~Wittig, 
\emph{JHEP} {\bf 0404}, 013 (2004) 

\bibitem{LMA2} 
T.A.~DeGrand and S.~Schaefer, 
\emph{Comput. Phys. Commun.} {\bf 159}, 185 (2004) 
[{\tt hep-lat/0401011}].

\bibitem{ata} 
J.~Foley {\it et. al}. (TrinLat collaboration), 
\emph{Phys. Commun} {\bf 172}, 145 (2005)
[{\tt hep-lat/0505023}].

\bibitem{fixedQ}
H.~Fukaya {\it et al.} (JLQCD collaboration),
\emph{Phys. Rev. D} {\bf 74}, 094505 (2006)
[{\tt hep-lat/0607020}].

\bibitem{Nf2:JLQCD}
S.~Aoki {\it et al.} (JLQCD collaboration),
\emph{Phys. Rev. D} {\bf 78}, 014508 (2008)
[{\tt 0803.3197[hep-lat]}].


\bibitem{m_s}
J. ~Noaki {\it et. al}. (JLQCD collaboration),
in preparation.

\bibitem{HBChPT}
A.~Walker-Loud, 
\emph{Nucl. Phys. A} {\bf 747}, 476 (2005) 
[{\tt hep-lat/0405007}].
 

\bibitem{m_N}
C. ~Amsler,{\it et al}. (Particle Data Group).
 \emph{Phys. lett. B} {\bf 667}, 1 (2008).

\bibitem{Fukugita} 
M.~Fukugita et al., 
\emph{Phys. Rev. D} {\bf 51}, 5319 (1995) 
[{\tt hep-lat/9408002}].

\bibitem{Dong} 
S.J.~Dong, J.F.~Lagae and K.F.~Liu, 
\emph{Phys. Rev. D} {\bf 54}, 5496 (1996) 
[{\tt hep-ph/9602259}].

\bibitem{SESAM} 
S.~Gusken et al. (SESAM collaboration),
\emph{Phys. Rev. D } {\bf 59}, 054504 (1999)
[{\tt hep-lat/9809066}].

\bibitem{UKQCD} 
C.~Michael, C.~McNeile and D.~Hepburn (UKQCD collaboration), 
\emph{Nucl.Phys.Proc.Suppl.} {\bf 106}, 293 (2002)
[{\tt hep-lat/0109028}].

\bibitem{ohki:lat09} 
H.~Ohki {\it et al.} (JLQCD collaboration),
in these proceedings.


\end{thebibliography}
\end{document}